\documentclass[apl,aip,reprint,]{revtex4-1}
 \newcommand{\beq}{\begin{equation}}
\newcommand{\eeq}{\end{equation}}
\usepackage{graphics}
\usepackage{epsf}
\usepackage{amsmath}
\usepackage{amsfonts}
\usepackage{amssymb}
\usepackage{graphicx}
\usepackage{epsfig}

\begin{document}

\title{ Epitaxial growth of (111)-oriented LaAlO$_3$/LaNiO$_3$  ultra-thin superlattices  }
\author {S. Middey}
\email{smiddey@uark.edu  }
\affiliation  {Department of Physics, University of Arkansas, Fayetteville, Arkansas 72701, USA}
\author{D. Meyers}
\affiliation  {Department of Physics, University of Arkansas, Fayetteville, Arkansas 72701, USA}
\author{M. Kareev}
\affiliation  {Department of Physics, University of Arkansas, Fayetteville, Arkansas 72701, USA}
\author{E. J. Moon}
\affiliation  {Department of Physics, University of Arkansas, Fayetteville, Arkansas 72701, USA}
\author{B. A. Gray}
\affiliation  {Department of Physics, University of Arkansas, Fayetteville, Arkansas 72701, USA}
\author{X. Liu}
\affiliation  {Department of Physics, University of Arkansas, Fayetteville, Arkansas 72701, USA}
\author{J. W. Freeland}
\affiliation {Advanced Photon Source, Argonne National Laboratory, Argonne, Illinois 60439, USA}
\author{ J. Chakhalian}
\affiliation  {Department of Physics, University of Arkansas, Fayetteville, Arkansas 72701, USA}


\begin{abstract}

The epitaxial stabilization of  a single layer or superlattice structures composed of complex oxide materials on polar (111) surfaces is severely burdened by   reconstructions at the interface, that commonly arise  to neutralize the polarity.  We report on the  synthesis of high quality LaNiO$_3$/mLaAlO$_3$  pseudo cubic (111) superlattices on polar (111)-oriented LaAlO$_3$,   the proposed complex oxide candidate  for   a topological insulating behavior. Comprehensive X-Ray  diffraction   measurements, RHEED,   and  element specific resonant X-ray absorption spectroscopy  affirm their high  structural  and chemical quality.  The study offers an opportunity to fabricate interesting interface and topology controlled  (111) oriented superlattices based  on ortho-nickelates.
\end{abstract}

\maketitle

Over the past few years, complex oxide superlattices (SL) with correlated carriers   have been widely studied  owing to the range of  exciting phenomena emerging at the interface which are unattainable in  the bulk constituents.~\cite{jak_nm,tokura_nm_12}  Recently, active experimental investigations~\cite{lno_jian_apl,lno_jian_prb,lno_epl,lno_keimer_science,lno_keimer_nm} on the class of SLs consisting of paramagnetic metal LaNiO$_3$ (LNO) and   LaAlO$_3$ (LAO) were initiated   after the prediction of a possible high $T_c$ superconductivity   in the LaNiO$_3$/La$M$O$_3$ heterostructures (where La$M$O$_3$ is a wide band-gap insulator).~\cite{ln_th1,ln_th2} The experimental realization   of  LNO/LAO SLs grown on a (001)~\cite{notation} surface of SrTiO$_3$ (STO), however, revealed the presence of an unexpected transition to Mott insulating ground  state with  antiferromagnetic  order due to  quantum confinement  and the effect of $d$-orbital polarization by the interface.~\cite{lno_jian_prb,lno_keimer_science}   Inspired by this approach, several recent theory  proposals   have been put forward  regarding the physics which  may  emerge in  a bilayer of LaNiO$_3$ sandwiched between LaAlO$_3$  layers  grown along the [111] crystallographic direction. Specifically, the theory predicts  the appearance of  exotic  topological phases ($e.g.$ Dirac half-semimetal phase,  quantum anomalous Hall insulator phase or   ferromagnetic nematic phase) modulated by  the strength of electron-electron correlations.~\cite{satoshi_prb, lno1_fiete,lno_fiete, nagaosa_nc}   To date,  very  little experimental work  have been done  to develop such heterojunctions along the [111] direction to verify  the theoretical  predictions about  this   class of artificial    materials with interesting electronic  and magnetic ground states.

One of the main challenges in developing growth along  [111]  is  that commonly used substrates such  as LaAlO$_3$ (or SrTiO$_3$) consist of alternating LaO$_3^{3-}$Al$^{3+}$ (or SrO$_3^{4-}$Ti$^{4+}$) charged planes stacked  along the  [111] direction. The epitaxial thin film growth  along this highly polar direction~\cite{lcfo_ben,crre_apl,feta_apl} is  far less understood due to the possible occurrence of  complex surface reconstructions  that act to compensate for  the polar mismatch.   For example,  recently it  has been demonstrated that for the systems with  strong polarity mismatch e.g. BiFeO$_3$ on STO or CaTiO$_3$ on LAO, the epitaxial stabilization  is possible only if a ``screening''  buffer layer is grown first on the polar surface.~\cite{polar_apl} On the other hand, the polarity matching at the interface can have strong influence on the epitaxial growth, defects formation and overall stoichiometry of the layers as observed by the marked interfacial electronic reconstruction for polar LNO film grown on the top of charge neutral STO vs.  polar LAO (001) surface.~\cite{jian_apl} 

In this letter, we present the results of  artificial layer-by-layer growth of an unique class of (111)-oriented 2LNO/$m$LAO  heterostructures (with $m$ = 2, 3, and 4 unit cells)  on  LAO (111)  single  crystal (see Fig. 1(a)).  The LAO substrate was selected to eliminate the  effects of lattice  mismatch (i.e. strain) between the layers, which  otherwise may hinder the quality of  growth. The extensive characterization  using  reflection high energy electron diffraction (RHEED), atomic force microscopy (AFM), X-ray diffraction (XRD), and synchrotron based resonant X-ray absorption spectroscopy (XAS) confirm  the high structural, chemical and electronic quality of these superlattices  designed to facilitate the realization of the   geometry driven electronic and magnetic phases.  

 \begin{figure}
\begin{center}
\resizebox{8.3cm}{!}
{\includegraphics*[150pt,360pt][405pt,750pt]{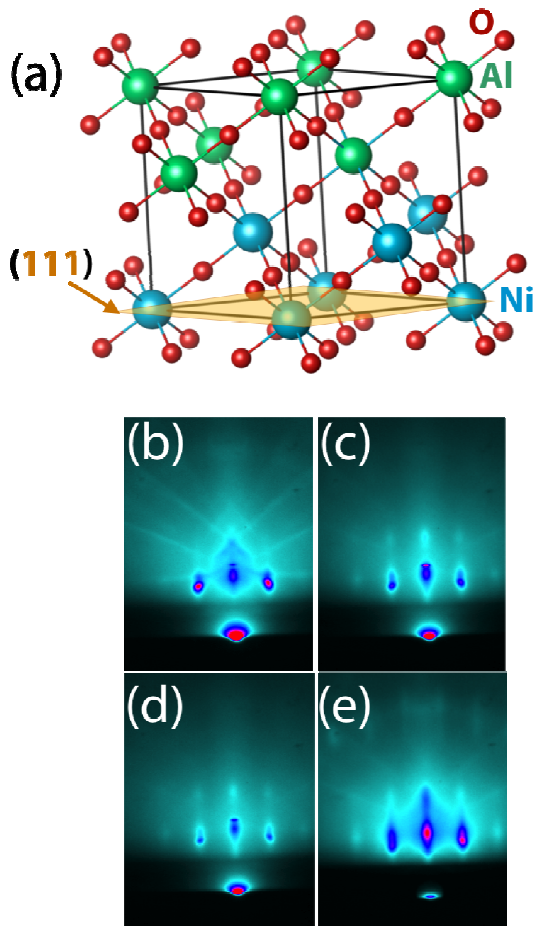}} \\
\caption{(color online) (a) Structural  arrangement  of  atoms in a bilayer of  LaNiO$_3$/LaAlO$_3$  viewed along (111). La atoms are omitted for  clarity. Note,  the RHEED images are taken along [1 -1 0] on (b) LAO substrate; (c)-(d) during the growths; (e) on 2LNO/2LAO film after cooling to room temperature.}
\end{center}
\end{figure}

 Fully epitaxial 2LNO/\textit{m}LAO SLs were grown by laser MBE operating  in   interval deposition mode on commercially available high-quality mixed terminated LAO (111) substrates.~\cite{misha_jap} The in-situ growth was monitored  by   high-pressure RHEED. The growth  was carried out under 50 mTorr of partial pressure of oxygen at a  deposition rate of 20-30 Hz; the substrate  temperature was set at 670$^\circ$C. To maintain correct oxygen stoichiometry the grown samples were subsequently post annealed  \textit{in-situ} for 30 minutes in  1 Atm of ultra pure oxygen.  Electrical d.c. transport was  performed in a commercial physical properties measurement system using the  {\it van der Paw} geometry.

\begin{figure}
\begin{center}
\resizebox{8.0cm}{!}
{\includegraphics*[0pt,0pt][250pt,230pt]{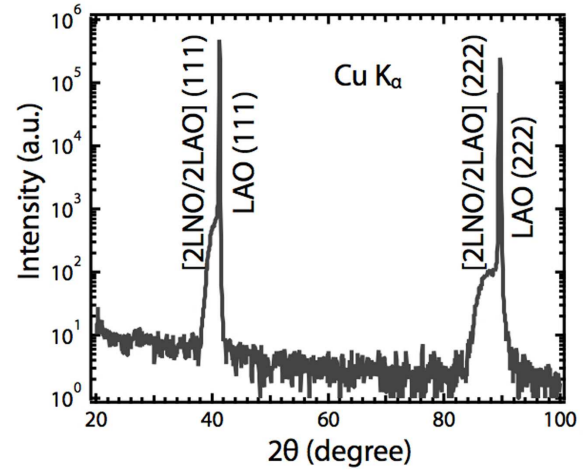}} \\
\caption{(color online)  2$\theta$-$\omega$ XRD scan for [2LNO/2LAO] superlattice using Cu K$_\alpha$ radiation.}
\end{center}
\end{figure}

In order to elucidate  how the formally  polar (111) surfaces of LAO substrate neutralizes the charge, we have investigated the as-received LAO\ substrate by combination of atomic force microscopy (AFM) and   X-ray photoelectron spectroscopy  (XPS) obtained at the different core states at  both the  grazing  and normal orientation between the detector and the   sample surface.  The detailed characterization has revealed  that the substrate possesses mixed termination ($i.e.$ Al$^{3+}$ and (LaO$_3$)$^{3-}$) and the charge polarity is largely compensated by the formation of    hydroxides on the LAO surface~\cite{supporting}; heating of the LAO\ substrate to the high growth temperature removes  hydroxides.~\cite{laoh} In addition, a sequence of RHEED images was recorded at different  stages of the growth and are  shown in Fig. 1(b)-(e).  As seen in Fig. 1(b), the   streak pattern observed along the Laue circles and the strongly  developed  Kikuchi lines  clearly exclude  the possibility of faceted morphology of the bare LAO (111) substrate~\cite{rheedbook}. Furthermore, as seen 
in   Fig. 1(c) and (d), the specular intensity  oscillations (not  shown) along  with the   retention of the original (111) diffraction pattern  after the  subsequent deposition of  LNO and LAO  layers     confirm the    layer-by-layer growth along [111] direction; upon cooling  down to  ambient temperature,  the high morphological quality of the [2LNO/2LAO] superlattices remained unaltered (see  Fig. 1(e)).  In order to gain further insight in to the structural quality of growth along the [111] direction, we have recorded X-ray diffraction $2\theta-\omega$ pattern using Cu K$_\alpha$ radiation. Fig. 2 displays the X-ray diffraction  obtained in the vicinity of the LAO (111) and (222) reflections. The  XRD data  further support  the proper [111] orientation of the heterostructures and  negate the possibility of a detectable impurity phase formation. The analysis of the Bragg reflections yields  the expected  1.5\%  tetragonal enhancement of the unit cell along the growth direction  due to the in-plane compressive strain characteristic of   full heteroepitaxy.

 \begin{figure}  
\begin{center}
\resizebox{8.0cm}{!}
{\includegraphics*[0pt,0pt][240pt,300pt]{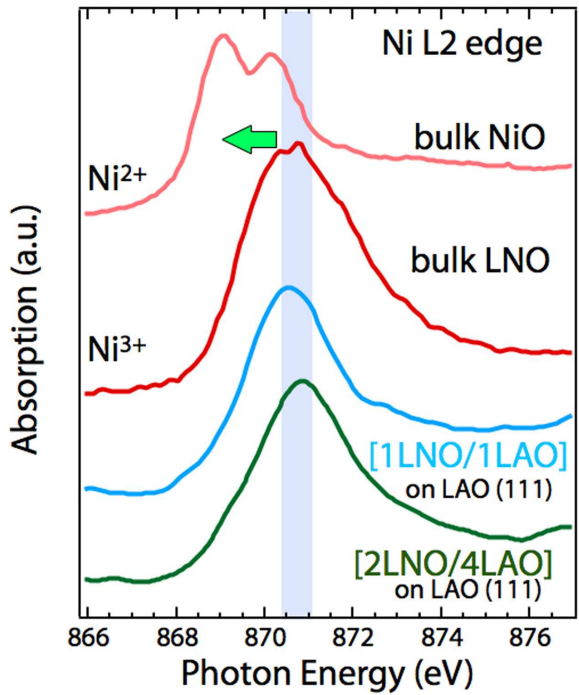}} \\
\caption{(color online)Polarization averaged Ni $L_2$ XAS spectra  obtained for  1LNO/1LAO and 2LNO/4LAO (111) superlattices and  compared with bulk Ni(2+)O and bulk LaNi(3+)O$_3$ reference samples.}
\end{center}
\end{figure}

Since perovskite rare-earth nickelates are characterized by  the unusually  high valence state of Ni$^{3+}$, the high  structural quality of the ultra-thin SLs  cannot  guarantee their proper stoichiometry, especially  against the formation of  oxygen  defects. In order to investigate the electronic and chemical structure of  the SLs,  measurements on the Ni L-edge, sensitive to  the Ni  charge state were performed in both total electron yield (TEY) and total fluorescence yield (TFY) modes at the soft X-ray branch of the 4-ID-C beam line at the Advanced Photon Source in Argonne National Laboratory. The   absorption data are depicted in Fig. 3 along with the absorption on the    bulk  LaNi(3+)O$_{3}$  and   Ni(2+)O reference samples taken in TFY mode to  avoid the chemical issues associated with the surface. As clearly seen, the absence of  the  characteristic for Ni$^{2+}$ multiplet  structure in the  line-shape of Ni L$_{2}$-edge and the  peak position  at 870.5 eV perfectly aligned to  the bulk LaNi(3+)O$_3$ reference  testify for the expected 3+ charge state of Ni and further affirms the lack of any sizable charge transfer at the LNO/LAO interface. These results validate a proper superlattice  stoichiometry critical for   material quality which is  in sharp  contrast to the case of polar LNO grown on non-polar STO (100) surface.~\cite{jian_apl}

 \begin{figure}  
 \begin{center}
\resizebox{8.0cm}{!}
{\includegraphics*[0pt,0pt][240pt,295pt]{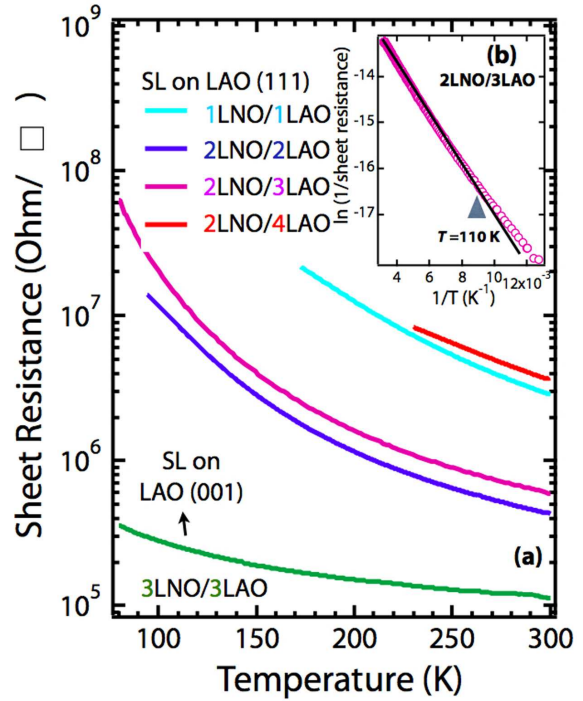}} \\
\caption{(color online)(a) Sheet resistance for different superlattices are plotted as a function on  temperature.  (b) The conductance can be fitted by activated behavior above 110 K.}
\end{center}
\end{figure}

After   structural and local chemical quality  had been  verified we proceed with the macroscopic d.c. transport measurements.  Figure 4(a) shows the  evolution of sheet resistance with temperature for a series of SL with varying LAO layer  thickness. The data shows a strong variance with the metallic state deduced from the earlier band structure calculations~\cite{lno_fiete};  all the heterostructures are highly insulating instead, hinting for a possible Mott behavior.  However, further work are needed  to confirm the nature of this insulating phase.  The analysis of the transport temperature dependence under assumption of  a single exponentially activated gap [$\sigma\propto$ exp(-$E_g/2k_BT$)] illustrated in Fig. 4(b) yields the value of  gap $E_g$ of $\sim 95$ meV.  The importance of  the geometrical  stacking  of  atoms is  further highlighted by the observation  that the  (111) SL's  exhibit  markedly higher resistivity  compared to (001) SL (see Fig. 4(a)).  Moreover, the transport for the 3LNO/3LAO SL grown on LAO (001) follows the  variable-range hopping  ~\cite{lno_jian_prb} contrary to the single gap activated behavior observed in the (111) SLs.

In summary, we have synthesized  the high-quality fully epitaxial 2LNO/\textit{m}LAO superlattices on formally  polar LAO (111).  A combination of RHEED,  XRD and XAS studies have confirmed   the excellent structural, chemical  and electronic quality of these heterostructures. All the fabricated SL's are insulating in nature,  contrary to the  proposed from \textit{ab-initio} calculations metallic ground state. The presented results also open the exciting prospects for fabrication of other ortho-nickelate based (111) superlattices $(e.g.$ NdNiO$_3$/LAO and  EuNiO$_3$/LAO) with  interesting geometry-driven magnetic and electronic  ground states.


The authors acknowledge fruitful discussions with G. A. Fiete and D. Khomskii.  J. C. was supported by DOD-ARO under the Grant No. 0402-17291 and NSF Grant No. DMR-0747808. Work at the Advanced Photon Source, Argonne is supported by the U.S. Department of Energy, Office of Science under Grant No. DEAC02-06CH11357.







\begin{thebibliography}{99}
\bibitem{notation} All of the crystallographic directions, plane are in psedocubic notations.
\end{thebibliography}
\end{document}